\documentclass[a4paper]{jpconf}
\usepackage{indentfirst}
\usepackage{graphicx}
\usepackage{amsmath}
\usepackage{eso-pic}
\usepackage{lineno}
\usepackage{tabto}
\usepackage{tikz}

\usepackage[sorting=none]{biblatex}
\newcommand{\comment}[1]{}

\begin{document}

\comment{
    \AddToShipoutPictureBG{
    \begin{tikzpicture}[remember picture, overlay,
                        help lines/.append style={
                            line width=0.05pt, color=blue!10},
                        major divisions/.style={
                            help lines,line width=0.1pt, color=red!20} 
                        ]
        \draw[help lines](current page.south west) grid[step=10pt]
                        (current page.north east);
        \draw[major divisions](current page.south west) grid[step=50pt]
                        (current page.north east);
    \end{tikzpicture}
    }
}

\title{Multichannel SiPM test readout system for gamma ray
measurements with monolithic inorganic CeBr$_3$}

\author{Veronika Asova$^1$, Galin Bistrev$^2$, Valentin Buchakchiev, Venelin Kozhuharov}
\address{Faculty of Physics, Sofia University, 5 J. Bourchier Blvd., 1164 Sofia, BG}
\ead{$^1$viasova@uni-sofia.bg, $^2$gbistrev@uni-sofia.bg}

\begin{abstract}
    Energy resolution and the detection efficiency 
    for gamma quanta are fundamental properties in the
    construction of detectors for ionizing radiation.
    In this study, a SiPM-based photodetector coupled to 
    a monolithic inorganic CeBr$_3$ crystal is exposed to
    gamma rays in order to study the performance of
    the CeBr$_3$ crystal. Measurements are made using
    three different radioactive sources - $^{137}$Cs,
    $^{22}$Na and $^{60}$Co. For each source, the measurements
    are made using a few different values for the Bias
    voltage of the SiPM. Furthermore, two CeBr$_3$
    crystals with different thicknesses are used in order
    to study how detector efficiency is affected by crystal
    dimensions. A preliminary analysis of the data is
    presented.
\end{abstract}

\section{Introduction}
    Gamma ray spectroscopy is an important subject in many
    scientific areas. Applications for understanding of the 
    gamma ray energy spectrum include nuclear medicine, 
    environmental research, and cosmic studies, among many
    other fields. Conventional gamma ray detectors utilize
    materials capable of interacting with incoming gamma
    particles in order to produce a voltage signal proportional
    to their energies. In order to assess a gamma detector, we
    look at a number of key characteristics, including energy
    resolution, detection efficiency and the linearity of
    the output signal with variations in energy.
    
    In nuclear physics, precise time and energy measurements
    are essential. Utilizing a monolithic inorganic scintillator
    coupled to a photodetector provides a high detection
    efficiency due to the lack of dead zones from the optical
    insulation. Another advantage is a counting rate of up to
    $10^7$~counts/s and the ability to measure in a broad range
    of energies. Examples of crystals that see use due to their
    short decay times and high light yield are LaBr$_3$ and
    CeBr$_3$~\cite{bib:LaBrandCeBr}. LaBr$_3$ has a decay time
    of $\sim$25 ns and a light yield of $\sim$63000~ph/MeV
    \cite{bib:LaBr3Ce}. It is also useful for continuous
    calibration and monitoring of the detectors due to
    lanthanum's natural radioactivity. However, for energies
    below 1.5 MeV, this radioactivity becomes significant and
    can affect measurements~\cite{bib:china}. Therefore,
    CeBr$_3$ crystals are used in low-noise environments.
    
    CeBr$_3$ crystals have a decay time of $\sim$17 ns and a light
    yield of $\sim$68000~ph/MeV~\cite{bib:CeBr3prop},~\cite{bib:pdg}. 
    Currently, CeBr$_3$ crystals are being considered for
    applications in gamma ray astronomy such as Compton 
    telescopes~\cite{bib:applications}, nano-satellites 
    \cite{Toneva:2020hvl} and environmental monitoring 
    \cite{bib:Environment}. Therefore, gaining a better
    understanding of CeBr$_3$-based detectors and their
    capabilities is essential. Two sets of measurements are
    presented in the current article, using two CeBr$_3$
    based scintillation detectors coupled to a 144-channel
    SiPM-based photodetector. The experimental setup will
    be presented in the next section.

\section{Experimental setup}
\begin{figure}[!htp]
    \begin{center}
        \includegraphics[width=0.9\textwidth]{
            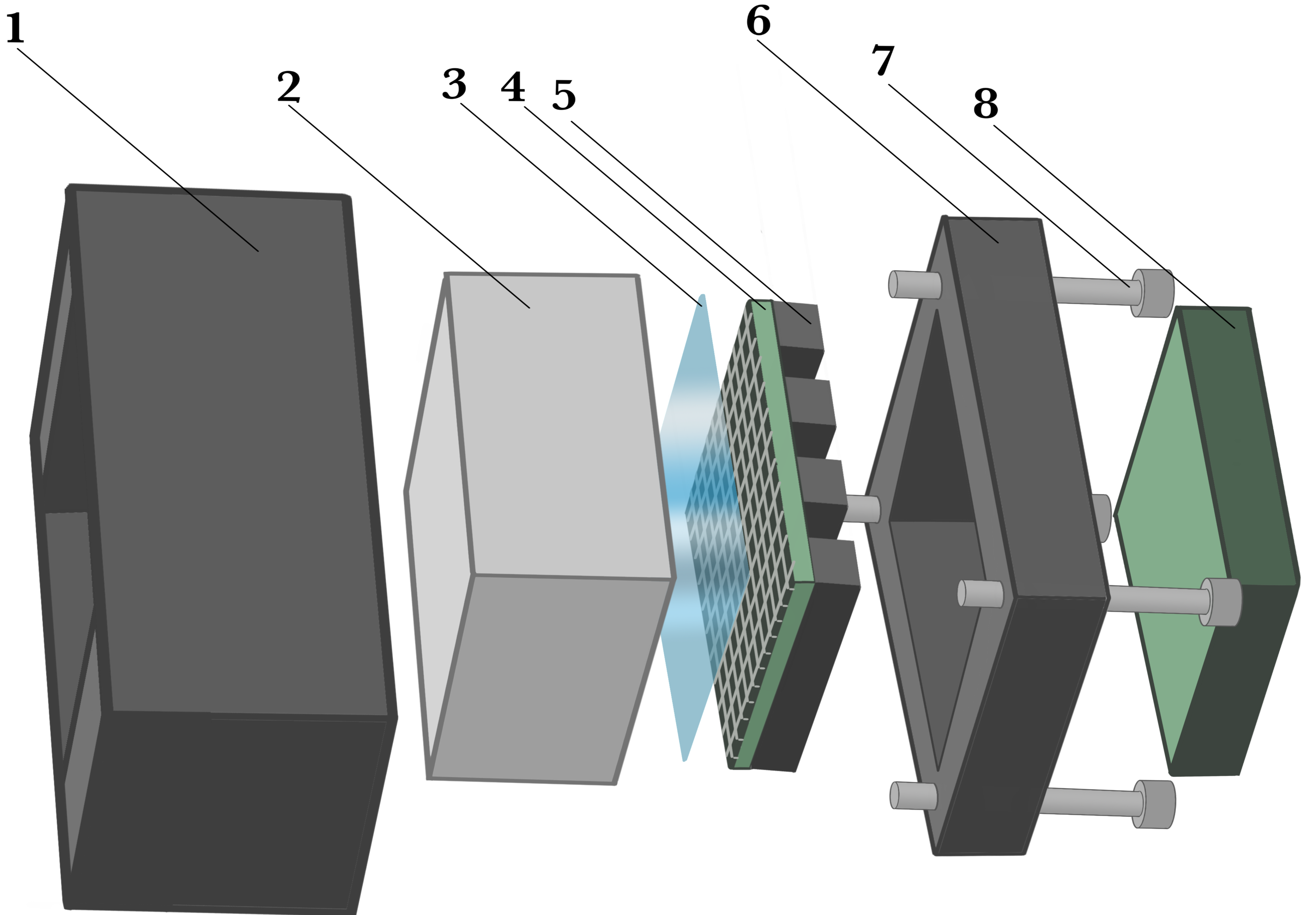
        }
        \caption{
            \label{fig:housing}
            A schematic representation of the SiPM+Crystal setup. 
        }
    \end{center}
    \end{figure}

    The setup incorporates an ONSEMI ARRAYC-30035-144PCB SiPM
    matrix consisting of 144 avalanche photodiodes (12x12 ARRAY
    of 3mm SMT sensors). A SiPM is chosen over a conventional
    photomultiplier tube (PMT) due to the SiPM's combination of high
    gain, high signal to noise ratio and low operational voltage
    in comparison to PMTs~\cite{bib:PMT}. The SiPMs are coupled
    to one of two CeBr$_3$ crystals with different dimensions - 
    51$\times$51$\times$25~mm$^3$, and 51$\times$51$\times$10~mm$^3$.
    Going forward, we will refer to the former as "thick crystal" and
    to the latter as "thin crystal". The SiPM matrix and the crystal
    are coupled via a 3D printed housing sealed with epoxy.
    The matrix and its housing are shown in Figure \ref{fig:housing},
    where the numbers point to:
    \begin{itemize}
        \item{ 1 - 3D printed upper casing }
        \item{ 2 - $CeBr_3$ scintillator crystal }
        \item{ 3 - silicone optical grease }
        \item{ 4 - SiPM matrix }
        \item{ 5 - Samtec 80-way connectors type QTE-040-03-F-D-A }
        \item{ 6 - 3D printed lower casing }
        \item{ 7 - Fasteners }
        \item{ 8 - AiT AB424T-ARRAY144P Tileable 4+24 Channel Hybrid
               Active Base }
    \end{itemize}

    \newpage
    Power is provided by an AiT ABPS power supply module, which also
    provides the biasing voltage necessary for the the operation of the
    SiPMs. Whenever a trigger is received, the output signals from each
    row and column are summed through the AiT AB424T carrier, thus
    producing 2$\times$12 output signals - $Ax_i$ and $Ay_i$. There
    are four final outputs $X^+$, $X^-$, $Y^+$, $Y^-$, which are 
    functions of $Ax_i$ and $Ay_i$:

    \comment{
        \begin{equation}
            \begin{gathered}
                X^+ = \sum_i{c_i * Ax_i} \\
                X^- = \sum_i{c_{12 - i} * Ax_i} \\
                Y^+ = \sum_i{c_i * Ay_i} \\
                Y^- = \sum_i{c_{12 - i} * Ay_i}
            \end{gathered}
        \end{equation}
    }

    \begin{equation}
        X^+ = \sum_i{c_i * Ax_i},~~~~~
        X^- = \sum_i{c_{12 - i} * Ax_i}
    \end{equation}

    \begin{equation}
        Y^+ = \sum_i{c_i * Ay_i},~~~~~
        Y^- = \sum_i{c_{12 - i} * Ay_i}
    \end{equation} 

    \noindent such that $c_i + c_{12-i} = 1.0833$, leading to 
    
    \begin{equation}
        {X^+} + {X^-}  = 1.0833 \times \sum_i{Ax_i}
    \end{equation}

    \begin{equation}
        {Y^+} + {Y^-}  = 1.0833 \times \sum_i{Ay_i}
    \end{equation}
    
\begin{figure}[!ht]
    \begin{center}
        \includegraphics[width=0.875\textwidth]{
            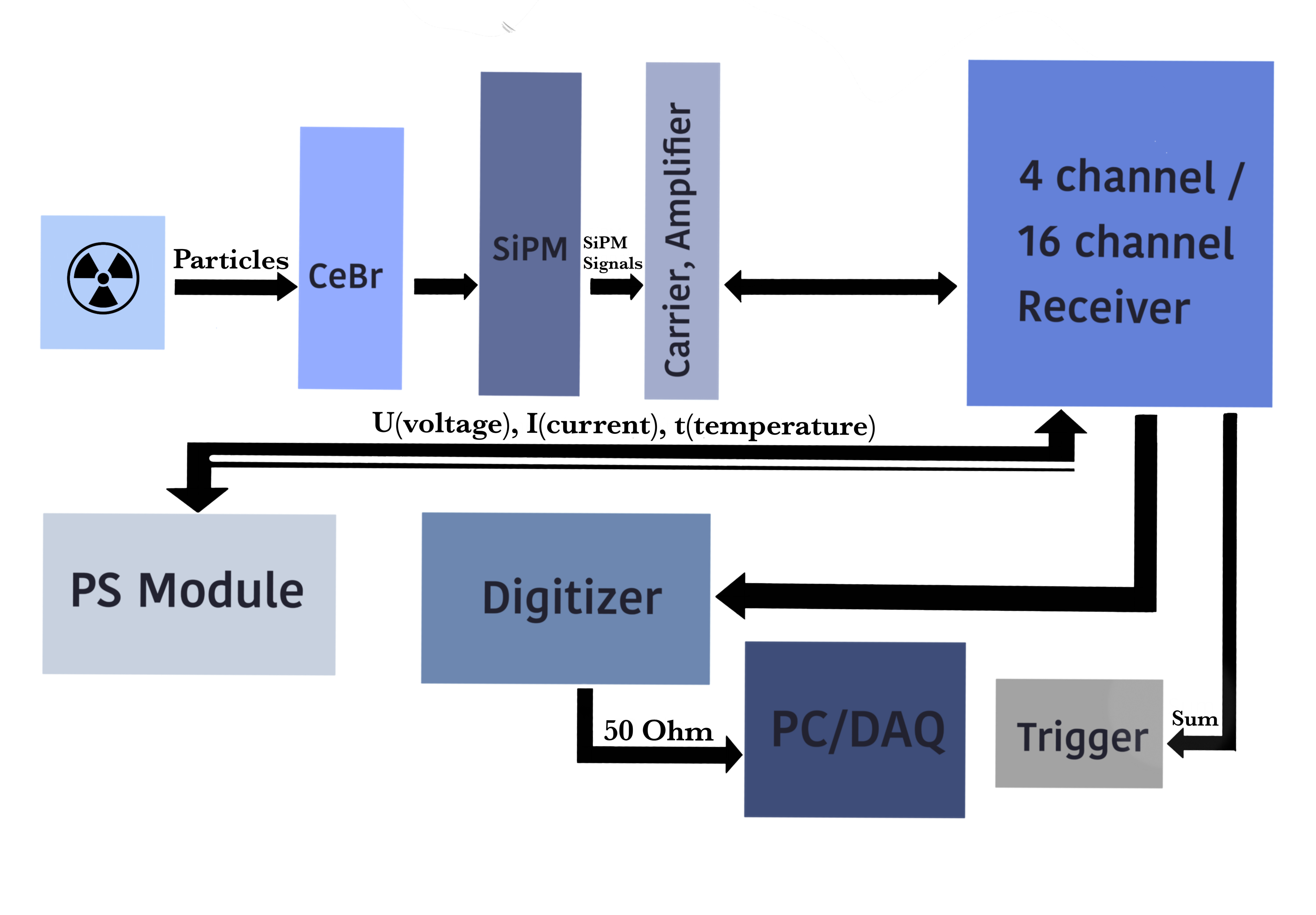
        }
        \caption{
            \label{fig:schematic}
            Schematics of the experimental setup which are
            also shown in \cite{bib:nafski2022}. 
            The radioactive source is one of the three available
            ($^{137}$Cs, $^{22}$Na, $^{60}$Co).
        }
    \end{center}
    \end{figure}

    Each set of signals is sent to either the 4-channel ABR4 or
    the 16-channel ABR16 receiver for amplification. The so-produced
    analog sum signals are then passed through a V1751 Digitizer
    and forwarded to a computer if over a set threshold (10~mV).
    The total number of samples is set to 1024 with a sampling rate
    of 10$^9$~Samples/s. A Data Acquisition (DAQ) software stores
    the data in a PC and prepares it for the subsequent analysis.
    A schematic representation of the full setup is shown on Figure
    \ref{fig:schematic}, while a more detailed description can be
    found in \cite{bib:nafski2022}.
    
    The measurements are made with collimated beams of gamma rays
    from the three radioactive sources ($^{137}$Cs, $^{22}$Na,
    $^{60}$Co). Collimation is achieved by using a pin-hole through
    a thick piece of Pb. For both the thick and the thin CeBr$_3$
    crystal we have data using three different bias voltages for the
    SiPMs (27.5~V, 28~V, 28.5~V). In the case of $^{137}$Cs, there
    is an additional measurement with 29~V bias.

\section{Results}
\begin{figure}[!ht]
    \begin{center}
        \includegraphics[width=0.8\textwidth]{
            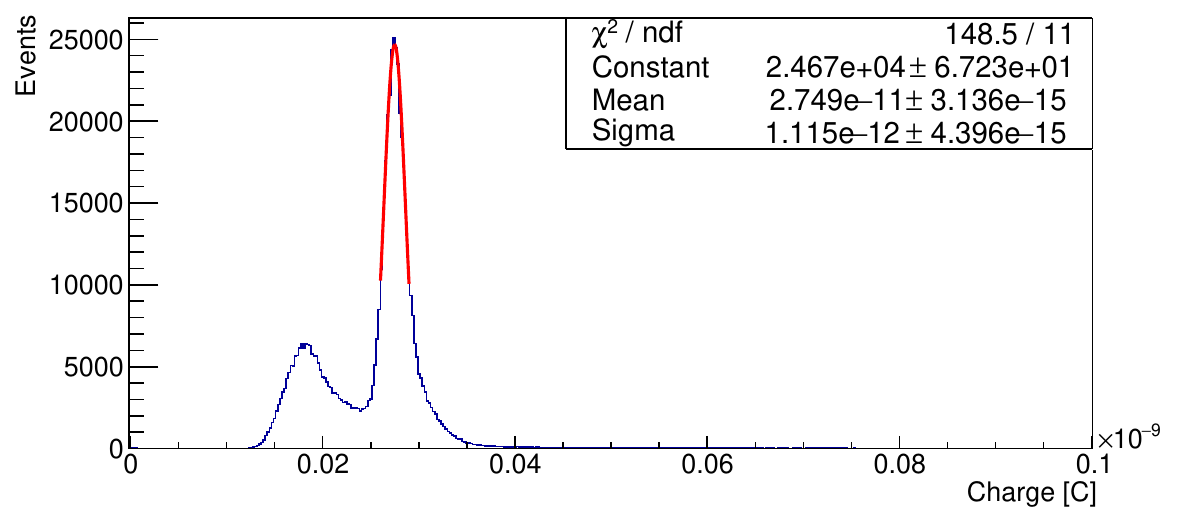
        }
        \caption{
            \label{fig:photopeak_27.5V}
            The spectrum for $^{137}$Cs with 27.5~V SiPM bias voltage.
            Fit function is gaussian.
        }
    \end{center}
    \end{figure}

\begin{figure}[!ht]
    \begin{center}
        \includegraphics[width=0.8\textwidth]{
            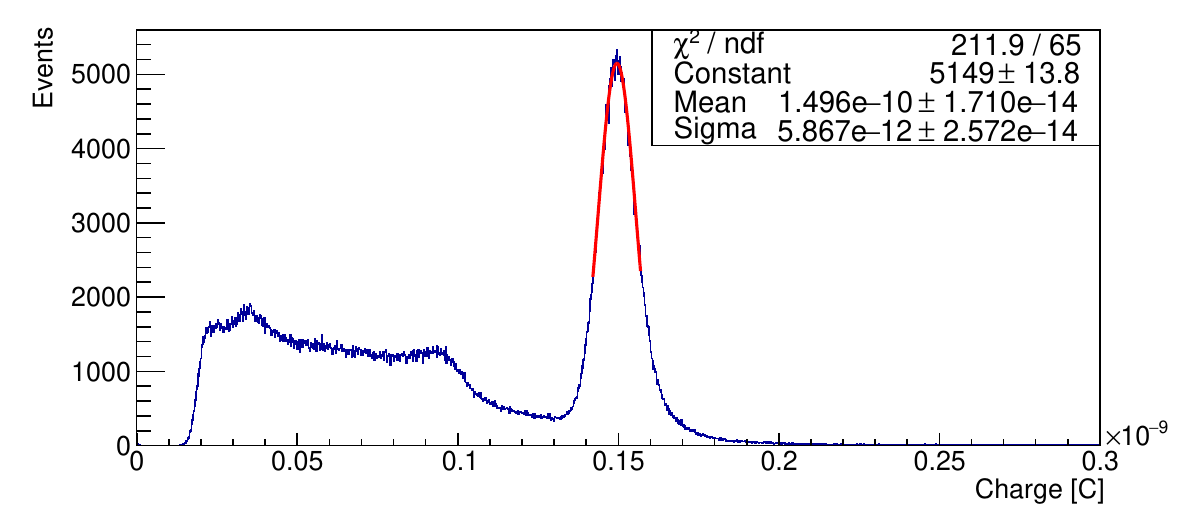
        }
        \caption{
            \label{fig:photopeak_29V}
            The spectrum for $^{137}$Cs with 29.0~V SiPM bias voltage.
            Fit function is gaussian.
        }
    \end{center}
    \end{figure}

    The total charge in each of the four channels was reconstructed
    using the formula:
    
    \begin{equation}
        Q=\sum_{i}^{n}\frac{U(t_i)}{R}\Delta{t},
    \end{equation}
    
    \noindent where $U(t_i)$ are the recorded amplitudes of the 
    signal in sample $i$, $\Delta{t}=1$~ns is the sampling step,
    and $R~=~50~\Omega$ is the input impedance of the digitizer.
    The recorded files were processed via a custom data analysis
    software package based on C/C++ and the ROOT framework~\cite{bib:ROOT}.
    On Figures~\ref{fig:photopeak_27.5V}~and~\ref{fig:photopeak_29V},
    the resulting spectra for $^{137}$Cs are presented for the SiPMs
    biased with 27.5~V and with 29~V.
    
    \comment{
        The fit function, that is used is:
    
        \begin{equation}
            f(x)=Ae^{(\frac{1}{2}\frac{(x-\mu)^2}{\sigma^2})}
        \end{equation}
    }

    \comment{
        Here A is a constant which sets the height of the peak,
        $\mu$ is the mean value of the energy of the peak and
        $\sigma$ is the standard deviation.
    }
        
\begin{figure}[!ht]
    \begin{center}
        \includegraphics[width=0.8\textwidth]{
            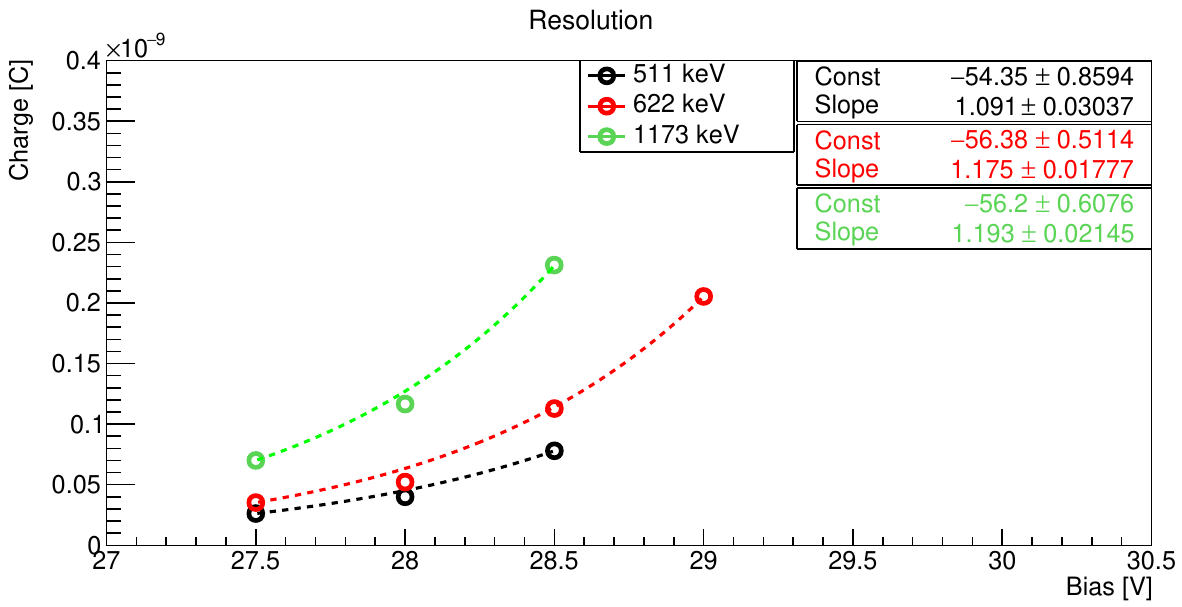
        }
        \includegraphics[width=0.8\textwidth]{
            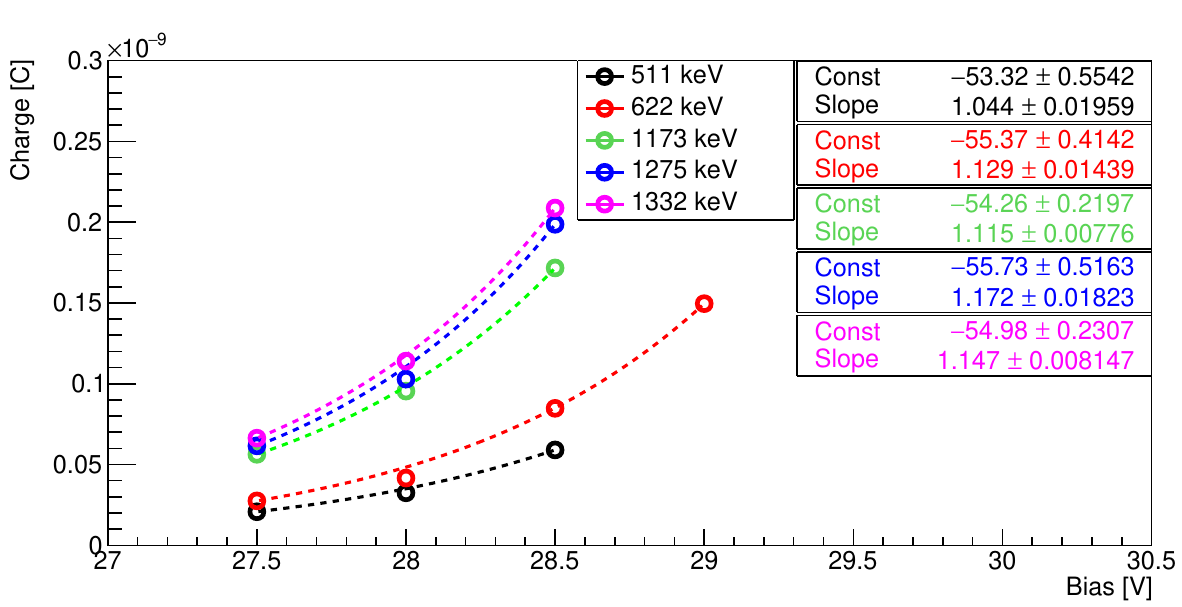
        }
        \caption{
            \label{fig:Charge}
            
        }
        Dependence of the reconstructed charge from Bias voltage for both
        the thin $CeBr_3$ crystal (top) and the thick one (bottom).
    \end{center}
    \end{figure}

    It can be seen that the photopeak of the radioactive isotope
    shifts to the right with increase in the bias voltage. This
    amplification is an intrinsic property of the SiPMs and
    should follow an exponential law~\cite{bib:exponential}.
    In Figure~\ref{fig:Charge} we show that this is indeed the
    case for our setup. The data points are fit using the function

    \begin{equation}
        f(x) = e^{(a+bx)}
    \end{equation}

    \noindent where the parameter \textit{a} is referred to as 
    "Const" in the graph legend and the parameter \textit{b}
    is referred to as "Slope".

    To determine how well the energy can be reconstructed from
    the charge, the dependence of the reconstructed charge on the
    impinging gamma energy is shown on Figure~\ref{fig:Calibration
    curve}. As detector response should be linear, the data points
    are fit to a line.
        
    \begin{figure}[!ht]
    \begin{center}
        \includegraphics[width=0.875\textwidth]{
            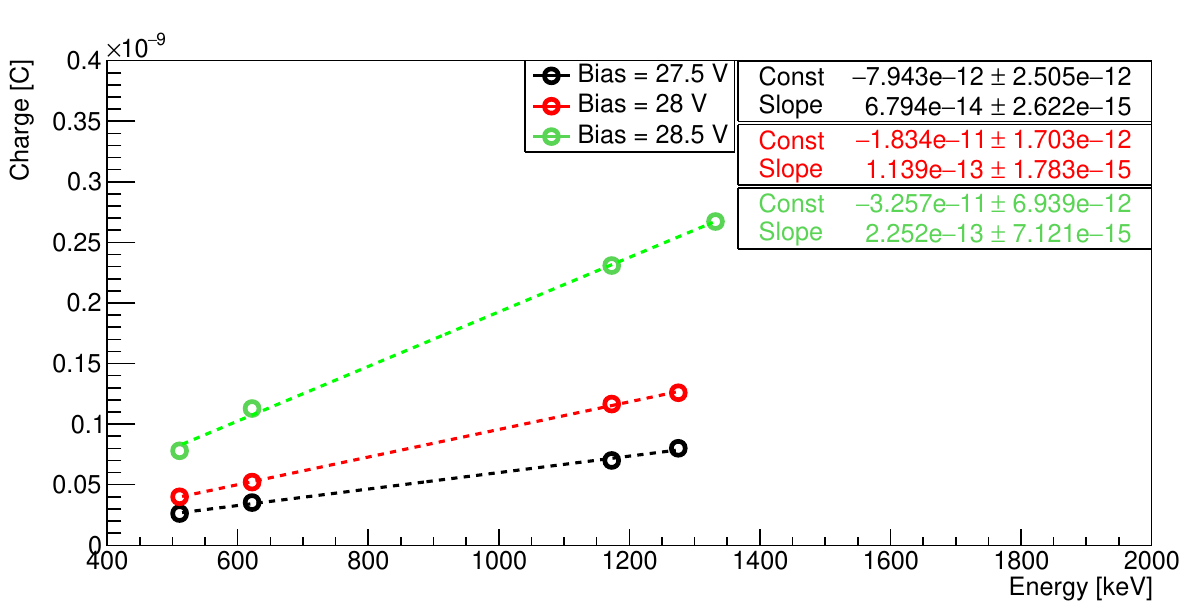
        }
        
        \includegraphics[width=0.875\textwidth]{
            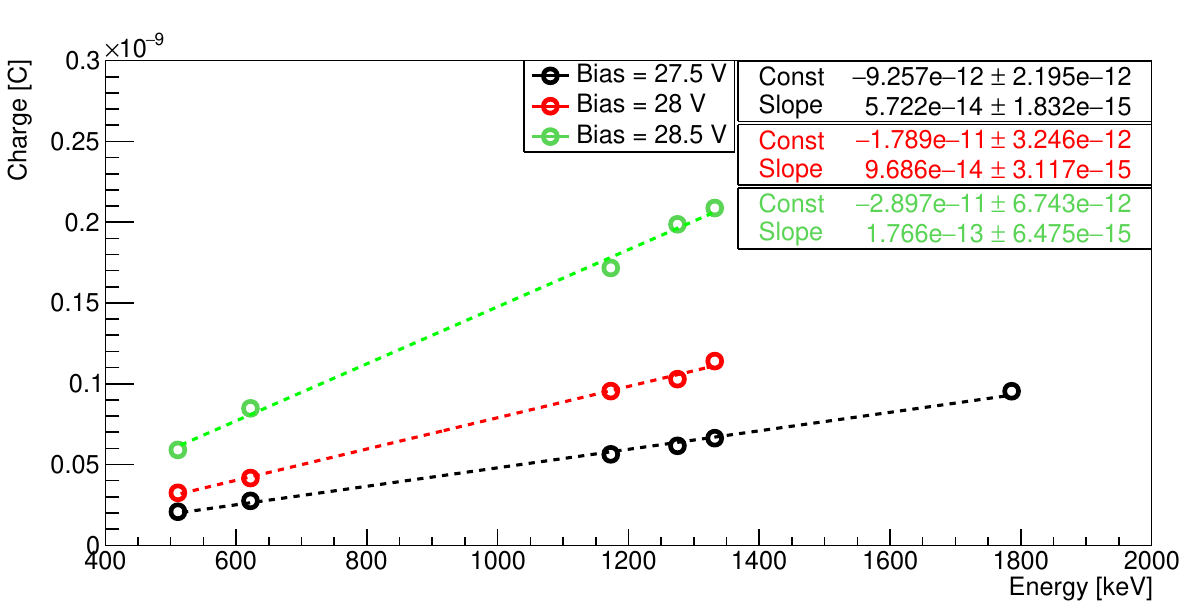
        }
        
        \caption{
            \label{fig:Calibration curve}
            Dependence of the reconstructed charge on the impinging
            gamma energy for both the thin $CeBr_3$ crystal (top)
            and the thick one (bottom).
        }
    \end{center}
    \end{figure}

    \newpage
    We calculate the detector's energy resolution via the formula:
    
    \begin{equation}
        \frac{\sigma(E)}{E} = \frac{\sigma(Q)}{\mu(Q)}
    \end{equation}
    
    where $\mu(Q)$ is the mean reconstructed charge value of the
    photopeak and $\sigma(Q)$ is its standard deviation.
    
    \begin{figure}[!htp]
    \begin{center}
        \includegraphics[width=0.9\textwidth]{
            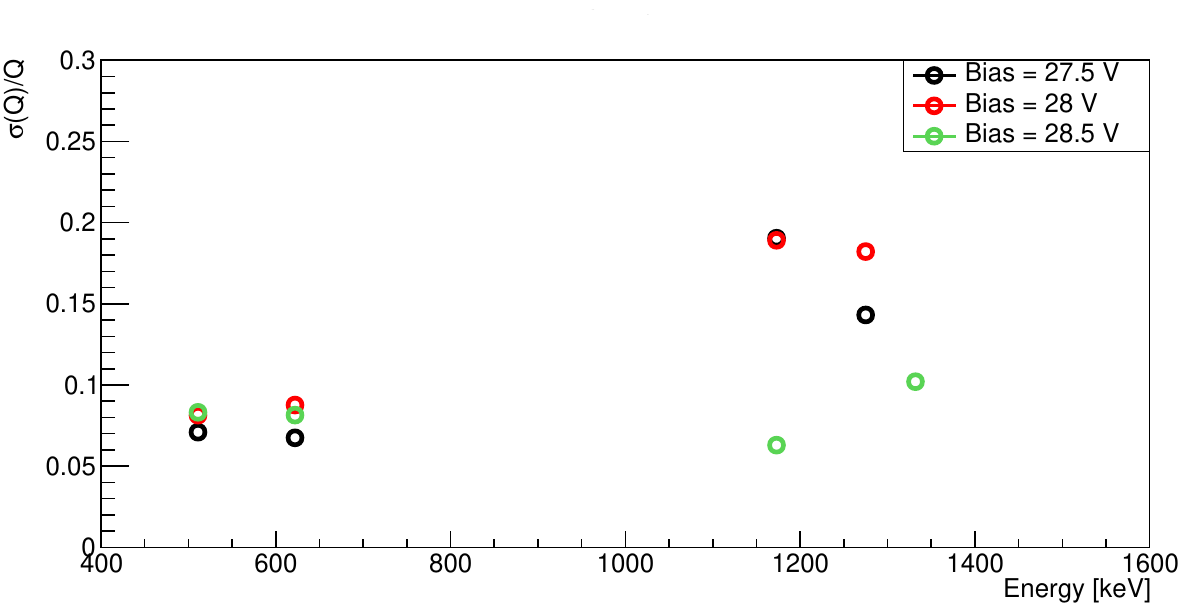
        }
        
        \includegraphics[width=0.9\textwidth]{
            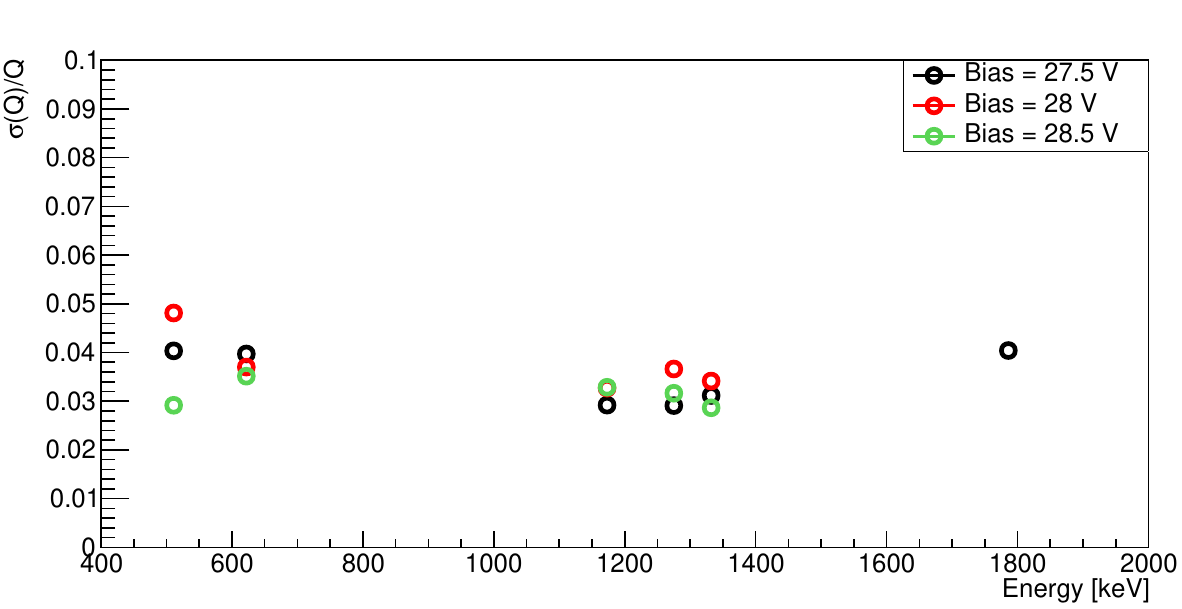
        }
        \caption{
            \label{fig:resolution}
            Energy resolution $\sigma_E/E$ as a function of
            impinging gamma energy for the thin crystal (top)
            and the thick crystal (bottom). 
        }
    \end{center}
    \end{figure}
    
    Figure~\ref{fig:resolution} shows that the energy resolution
    is at a plateau and that the value for this plateau differs
    depending on the crystal. For the thin $CeBr_3$ crystal, it is
    at $\frac{\sigma(E)}{E}~\approx~8\%$, while for the thicker one
    we have $\frac{\sigma(E)}{E}~\approx~4\%$. It can also be seen
    that there is no significant change with the SiPM bias voltage.
    All of this shows that the energy resolution of such a detector
    is directly affected by the properties of the crystal chosen. 

    \begin{figure}[!ht]
    \begin{center}
        \hspace{-10mm}
        \includegraphics[width=0.875\textwidth]{
            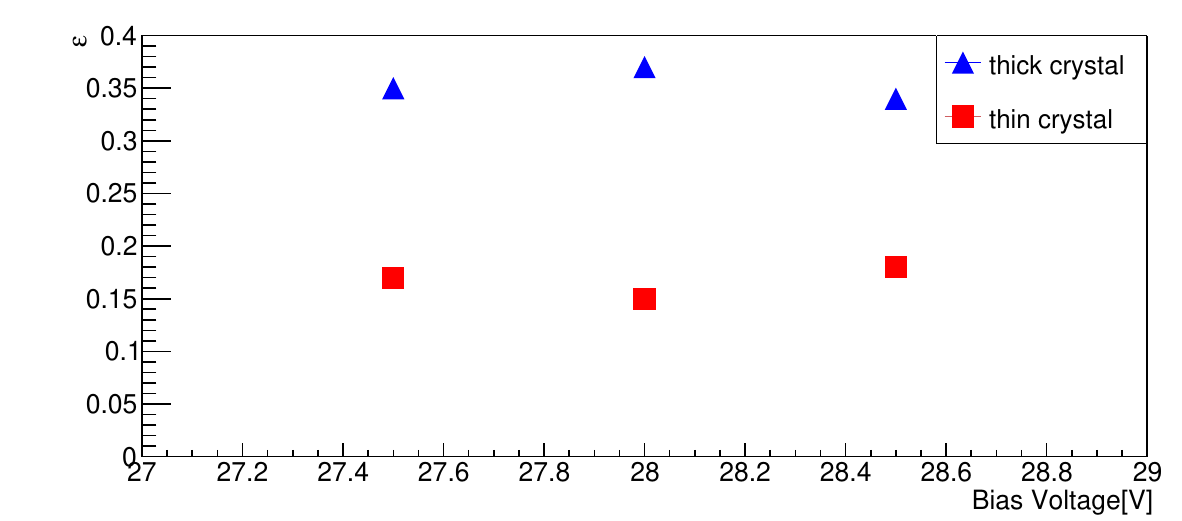
        }
        \caption{
            \label{fig:efficiency} The dependence of the relative
            efficiency of the detector on the bias voltage.
        }
    \end{center}
    \end{figure}

    \newpage
    On Figure~\ref{fig:efficiency} we show the dependence of what
    we call relative light yield efficiency of the detector from
    the bias voltage. We express it as 
    
    \begin{equation}
        \varepsilon = \frac{N(1275keV)}{N(511keV)}
    \end{equation}
    
    Here $N(1275~keV)$ and $N(511~keV)$ are the number of events that
    form the photopeaks of the individual energy lines of the $^{22}$Na
    isotope. We use the relative count for different lines of the same
    spectrum, due to lack of a reference source. Therefore we attempt to
    define the efficiency within a single run. Our expectation is a ratio
    of $\sim$1/2. Numbers below that would mean (an expected) loss of
    efficiency. Figure~\ref{fig:efficiency} shows that the relative
    efficiency of the thick crystal is $\varepsilon~\approx~0.35$, while
    the efficiency of thin one is $\varepsilon~\approx~0.15$. It remains
    mostly constant for the different bias voltages.

\section {Discussion}
    Based on the current data, it can be suggested that the energy 
    resolution of the thick crystal is marginally better than that of
    the thin crystal. The lack of sensitivity for the thin crystal also
    produces a less distinguishable photopeak, and a worse light yield
    efficiency as shown on Figure~\ref{fig:efficiency}. This can be
    attributed to the fact that fewer gamma quanta are being absorbed
    within the thin crystal than within the thick one. For the thinner
    crystal, a more conspicuous Compton edge could be observed. This is
    due to the Compton scattered photons leaving the scintillator without
    further interaction resulting in a larger energy deposition
    corresponding to a photon scattering angle at 180°. In the case of
    the thicker crystal, the Compton scattered gamma quanta would have a
    better chance of interacting again while inside the scintillator.
    
    \comment{
        , which is also the
        reason why the thin crystal has also a better resolution. There are
        fewer Compton scattering events in the thinner crystal since there
        are fewer gamma quanta interacting within its structure.
    }

    It can also be seen that the efficiency of the crystals remains
    mostly constant for different bias voltages, which means that the
    efficiency isn't dependant on parameters set by the electrical
    circuit, but is an intrinsic quality of the the crystal. The same
    statement can be made for the resolution, which also remains mostly
    constant for different bias voltages as shown on
    Figure~\ref{fig:resolution}.
    
    Comparing Figure~\ref{fig:photopeak_27.5V} and
    Figure~\ref{fig:photopeak_29V}, it can be see that the noise levels
    increase with the bias voltage. This can be attributed to thermal
    emission and dark current noise within the SiPMs. As seen on Figure~
    \ref{fig:Calibration curve}, both detector configurations exhibit
    similar linearity, which means both of them are capable of localizing
    energies of gamma quanta from different sources. However, in both
    cases there is no intersection with zero, which is a problem. A
    possible reason for this is an energy pedestal not being taken into
    account. Another possible cause may be an issue regarding a lack of 
    proportionality in the detector. The cause of this non-proportional
    response could be due to the fact that interaction of the gamma quanta
    with the environment leads to a non-uniform ionization density, which
    the light yield is also dependant on. This possibility is to be
    further looked into during future studies.

\section{Conclusions}
    The present research shows the advantages and disadvantages of two
    different configurations for a scintillating detector, using $CeBr_3$
    crystals of different thickness. Current results show that the thicker
    crystal ($51\times51\times25~mm^3$), due to a higher sum photon count,
    presents a better detection efficiency compared to the thin crystal
    ($51\times51\times10~mm^3$). Another result of this is that the
    thinner crystal experiences a relatively higher noise rate, due to
    the fewer interactions of gamma quanta within its structure. Finally,
    the thicker crystal also exhibits a better energy resolution.

    The relationship between the efficiency and the crystal thickness
    and how it compares to other crystals will be covered in more
    detailed subsequent studies. We would also tackle the subject of
    the detector's noted non-proportionality and whether or not this
    is an intrinsic quality of the detector. This is important, because
    currently the detector's response is not completely linear, as
    evidenced by the fact that the charge/energy line doesn't pass through
    zero. An energy pedestal not being accounted for is also one probable
    explanation for this phenomenon, but other possible reasons need
    to be explored and would be covered in future studies.

\section*{Acknowledgments}
    This work is partially supported by grant BG05M2OP001-1.001-0008,
    "National Centre on Mechatronics and Clean Technologies". In
    addition, VK and SL recognize that partially this study is
    financed by the European Union-NextGenerationEU, through the
    National Recovery and Resilience Plan of the Republic of Bulgaria,
    project SUMMIT BG-RRP-2.004-0008-C01.

\printbibliography[title={References}]

\end{document}